\documentclass[journal = nalefd,manuscript=letter,layout=twocolumn,articletitle = true]{achemso}  
\setkeys{acs}{articletitle = true}
\usepackage{graphicx}  
\usepackage{dcolumn}   
\usepackage{bm}        
\usepackage{amssymb}   
\usepackage{gensymb}
\usepackage{color}  

\usepackage{hyperref}
\usepackage{xcolor}
\usepackage{grffile} 

\title{Graphene whisperitronics: transducing whispering gallery modes into electronic transport}

\author{Boris Brun$^{1*}$, Viet-Hung Nguyen$^1$, Nicolas Moreau$^1$, Sowmya Somanchi$^{2,3}$,\\
Kenji Watanabe$^4$, Takashi Taniguchi$^5$, Jean-Christophe Charlier$^1$, Christoph Stampfer$^{2,3}$ \& Benoit Hackens$^{1*}$ }

\affiliation{
\normalsize{$^1$IMCN/NAPS \& MODL, Universit\'e catholique de Louvain (UCLouvain),}\\ 
\normalsize{B-1348 Louvain-la-Neuve, Belgium} \\
\normalsize{$^2$ JARA-FIT and 2nd Institute of Physics - RWTH Aachen University, Germany }\\
\normalsize{$^3$ Peter Gr\"unberg Institute (PGI-9), Forschungszentrum J\"ulich, 52425 J\"ulich, Germany}\\
\normalsize{$^4$ Research Center for Functional Materials, National Institute for Materials Science,} \\ 
\normalsize{1-1 Namiki, Tsukuba 305-0044, Japan.}\\
\normalsize{$^5$ International Center for Materials Nanoarchitectonics, National Institute for Materials Science,}\\
\normalsize{1-1 Namiki, Tsukuba 305-0044, Japan}\\
\normalsize{$^\ast$ To whom correspondence should be addressed;}\\
\normalsize{E-mails: boris.brun@uclouvain.be,benoit.hackens@uclouvain.be}
}

\date{}

\begin{document} 


\baselineskip24pt

\begin{abstract}
When confined in circular cavities, graphene relativistic charge carriers occupy whispering gallery modes (WGM) in analogy to classical acoustic and optical fields. The rich geometrical patterns of the WGM decorating the local density of states offer promising perspectives to devise new disruptive quantum devices. However, exploiting these highly sensitive resonances requires the transduction of the WGMs to the outside world through source and drain electrodes, a yet unreported configuration. Here we create a circular p-n island in a graphene device using a polarized scanning gate microscope tip, and probe the resulting WGMs signatures in in-plane electronic transport through the p-n island.
Combining tight-binding simulations and exact solution of the Dirac equation, we assign the measured device conductance features to WGMs, and demonstrate mode selectivity by displacing the p-n island with respect to a constriction.
This work therefore constitutes a proof of concept for graphene whisperitronics devices.
\end{abstract}


\maketitle

\subsection*{Introduction}
Whispering gallery modes (WGMs) denote the guided waves circulating along a concave surface, the seminal example of this phenomenon occurring in the whispering galleries of St Paul's cathedral with acoustic waves. 
This originally macroscopic-scale phenomenon found rich applications in nanotechnology with the ability to design resonating galleries, e.g. for electromagnetic waves, with nanometer-scale precision \cite{Vahala-2003,Foreman-2015}.
Taking advantage of the high quality factor and the reduced size of optical and mechanical whispering nano-resonators,
a new class of ultra-sensitive detectors has recently been developed, stretching the limits of physical and biological quantities detection down to a single molecule or virus \cite{Vollmer-2008,He-2011}. 
In the nano-electronic world, WGMs have proven successful to achieve distant coupling between GaAs quantum dots \cite{Nicoli-2018}. 
Recently, electronic WGMs have been reported in graphene \cite{Zhao-2015}, offering promising perspectives to develop a new class of  devices, taking advantage of the sensitivity of these geometrical resonances combined with the high technological integrability of graphene.

Graphene charge carriers behave as relativistic massless particles \cite{Novoselov-2005}, therefore sharing part of their fundamental properties with photons.
Consequently, graphene has early been proposed as an ideal platform to realize electron-optics experiments, in which p-n junctions play the role of lenses, prisms or fibers, and are used to guide, refract and transmit electrons and holes taking advantage of Klein tunneling \cite{Klein-1929,Cheianov-2007,Williams-2011,Rickhaus-2015,Liu-2017,Barnard-2017,Boggild-2017}.
As a seminal example of Dirac fermions optics, it has early been proposed \cite{Shytov-2008} and observed \cite{Young-2009} that Fabry-P\'erot (FP) interferences can arise between two facing p-n junctions, acting as semi-reflecting mirrors. When considering a circular p-n junction, FP interferences patterns then correspond to concentric rings, indeed visible in simulated wavefunctions \cite{Zhao-2015}. But FP interferences are only a sub-class of a larger set of WGMs : at larger angular momentum $m$, a manifold of patterns mixing radial and concentric symmetries emerge due to circular geometric symmetry, enriching the existing palette of Dirac fermions optics.

\begin{figure*}[h!]
\includegraphics[width = 1 \textwidth]{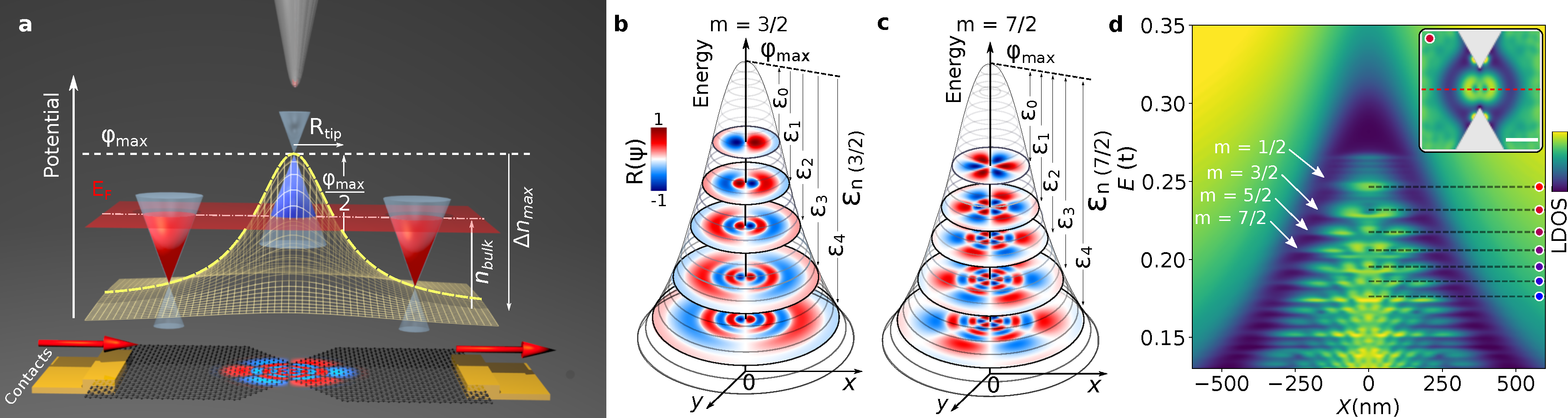}
\caption{\label{fig1} \textbf{Whisperitronics, transducing graphene whispering gallery modes to external contacts:}
\textbf{a}, Scheme of the electrostatic potential landscape: a biased AFM tip induces a Lorentzian potential whose maximum is $\varphi_{max}$ and half-width at half maximum is $R_{tip}$, centered on a graphene constriction doped at a Fermi energy $E_F$. Electronic transport is measured through the WGMs induced in the p-n island,the p region is below the tip. One WGM wave-function is plotted at the center with red and blue colors.
\textbf{b}, Real part of the wave-function $\Re (\psi_m (x,y))$ corresponding to the angular momentum $m$ = 3/2, at different resonant energies, calculated by solving the radial Dirac equation in presence of a Lorentzian potential, in an isotropic plane, without constriction. 
\textbf{c}, $\Re (\psi_m (x,y))$ for $m$ = 7/2, showing higher angular symmetry patterns.
\textbf{d}, LDOS in a p-n island, placed at the center of a 240 nm-wide graphene constriction,
as a function of energy $E$ in units of the hopping parameter \textit{t} and $x$ position along red dashed line in the inset. Inset: LDOS map for the third resonant mode. Scale bar: 200~nm.
}

\end{figure*}

Up to now, evidencing graphene p-n islands WGMs has relied on the use of a scanning tunneling microscope (STM)\cite{Zhao-2015, Ghahari-2017}, probing the local density of states (LDOS) through tunneling between the island and a metallic tip. However, the practical integration of WGMs-supporting p-n islands in future electronic devices will require to incorporate them in electron-optics setups, and combine them with other Dirac fermions optics tools\cite{Liu-2017,Barnard-2017,Boggild-2017}.
This transduction of the WGM devices to the outside world requires driving current between in-plane electrical contacts, a configuration in which WGMs have not been demonstrated in graphene to date.

Here we demonstrate the detection and manipulation of WGMs in in-plane transport \textit{through} a circular encapsulated graphene p-n junction, the essential ingredient to realize a new class of whisperitronics devices.
The p-n junction is created using the polarized tip of a scanning gate microscope (SGM) \cite{Eriksson-1996,Brun-PrB-2019}, 
which can be moved in the vicinity of a constriction  (Fig.~\ref{fig1}a). 
This demonstrates a surprising robustness of WGMs with respect to geometrical pinching of the circular island inside the constriction. Ultimately, WGMs selectivity is revealed by displacing laterally the p-n island with respect to the constriction, offering a very high degree of tunability. Consequently, our work paves the way towards controllable whisperitronics, merging these freshly revealed WGMs \cite{Zhao-2015} with the existing graphene electron-optics palette.

\subsection*{Results}

\begin{figure*}[h!]
\includegraphics[width = 1.0 \textwidth]{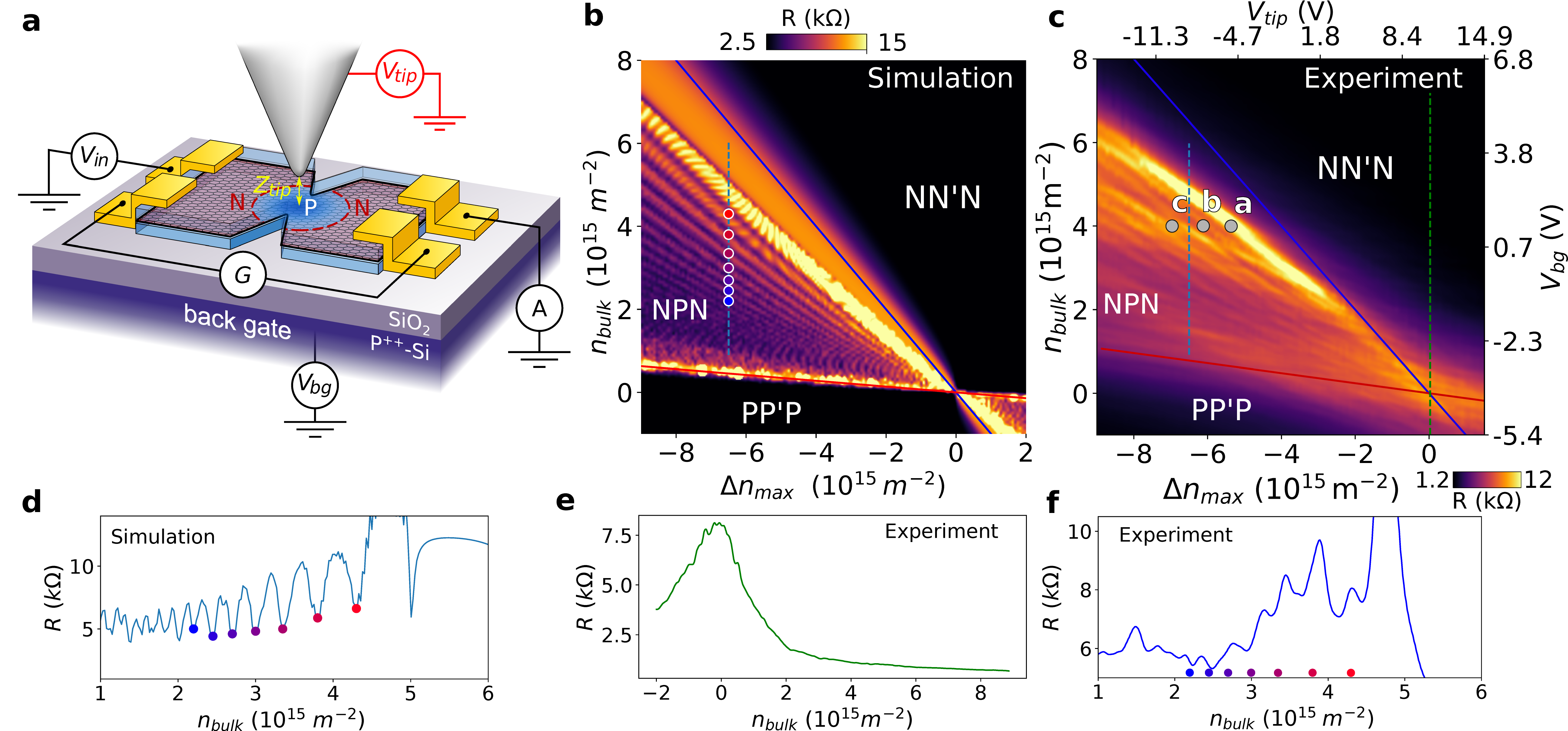}
\caption{\label{fig2} \textbf{Transport through WGMs in a constriction.}
\textbf{a}, Scheme of the experiment: a polarized AFM tip is placed above a constriction defined in encapsulated graphene. The tip potential is discussed in section 5 of the supplementary materials.
\textbf{b}, Calculated resistance as a function of the sheet carrier density $n_{bulk}$ and the maximum density change $\Delta n_{max}$, exhibiting oscillations in the N-P-N configuration. 
The value of $\Delta n_{max}$ = $-6.5 \times 10^{15} \rm m^{-2}$ used in Figs.~\ref{fig1}\textbf{d} and \ref{fig2}\textbf{d} is indicated by a blue dashed line as well as the 6 first resistance minima in this configuration that are labeled by colored dots (red to blue), reported in Figs.~\ref{fig1}\textbf{d} and \ref{fig2}\textbf{d} for comparison. N' and P' denote the carrier nature in the tip-pertubed region when it is similar to the sheet carrier type.
Blue line indicates the N-N'-N/N-P-N limit $n_{bulk}$ = $-\Delta n_{max}$. Note that it does not coincide with the resistance maximum.
\textbf{c}, Calculated resistance versus $n_{bulk}$, for $\Delta n_{max}$ = $-6.5 \times 10^{15} \rm m^{-2}$.
\textbf{d}, Resistance as a function of $V_{tip}$ and $V_{bg}$, recorded by placing the tip above the constriction at a distance $Z_{tip}$ = 90~nm from the graphene plane. 
Couples of $V_{tip}$ and $V_{bg}$ values for which the SGM images in Fig.~\ref{fig3}\textbf{a-c} are recorded are indicated by dots.
\textbf{e}, Constriction resistance as a function of $n_{bulk}$, for a neutral tip ($\Delta n_{max}$ = 0, green dashed line in c). \textbf{f},  Constriction resistance as a function of $n_{bulk}$, for $V_{tip}$ = -11~V ($\Delta n_{max}$ = $-6.5 \times 10^{15} \rm m^{-2}$) and $Z_{tip}$ = 90~nm, corresponding to blue dashed line in \textbf{c}. Colored dots correspond to minima in the simulated data (in Fig.~\ref{fig2}d).
}
\end{figure*}

Charge carriers in a graphene p-n island occupy a rich set of states \cite{Zhao-2015,Gutierrez-2016,Lee-2016,Jiang-2017}, labeled by different angular momentum \textit{m}, each of them exhibiting several resonant energies as exemplified in Fig.~\ref{fig1}b and \ref{fig1}c. To sense these different WGMs using in-plane electronic transport, the easiest way is to connect the resonator to two separate graphene regions by pinching the p-n island in a narrow channel, e.g. an etched constriction, as schemed in Figs.~\ref{fig1}a and~\ref{fig2}a.
Surprisingly, the WGMs are really robust, and survive when placing the p-n island in such a constriction. This remarkable fact is demonstrated in Fig.~\ref{fig1}d, where we plot the LDOS profile along transport axis in a circular p-n junction placed at the center of a 240~nm-wide constriction.
Despite the constriction, the pattern is strikingly similar to the unperturbed p-n island case (see section 3 of the supplementary materials). 
Noteworthy, the resonant states yield LDOS patterns that extend in the low density region (indicated by white arrows Fig.~\ref{fig1}d), as shown experimentally in STM in Refs.\cite{Ghahari-2017,Gutierrez-2018}, and supported by recent theoretical explorations \cite{Le-2020}.
The assignment of different \textit{m} values to these LDOS extensions is made possible by explicit comparison of the tight-binding simulations shown in Fig.~\ref{fig1}d (obtained using the Kwant code \cite{Groth-2014} with proper scaling \cite{Liu-2015}, as well as a home-made recursive green function code\cite{Nguyen-2010}), with the exact solution of the Dirac equation \cite{Nguyen-2018} (see section 2-3 of the supplementary materials).

The total resistance of the WGM resonator exhibits strong oscillations versus carrier density, as shown by Fig.~\ref{fig2}b and \ref{fig2}d. Remarkably, the resistance minima correspond to the extended WGMs, as indicated by the color dots showing the resonant energies of WGMs in Fig.~\ref{fig1}d, translated in carriers density and reported on the calculated resistance plots of Fig.~\ref{fig2}b and \ref{fig2}d. This is surprising, since WGMs arise due to internal reflections of the Dirac fermions on the border of the resonator.  On the other hand, they yield a non-zero LDOS in the low-density region at the edge, which favors a coupling with the Dirac continuum of states outside the resonator. These competing effects counter-intuitively result in a low total resistance. Finally, we show in Fig.~\ref{fig2}b the evolution of the resistance oscillations with $n_{bulk}$ and $\Delta n_{max}$, that correspond to the evolution of the different WGMs with energy and potential strength. Noticeably, these oscillations are not equally spaced in $n_{bulk}$, and their spacing increases as $\Delta n_{max}$ is made more negative. The location of minima also strongly depends on the geometry of the junction, as discussed in section S6 of the supplementary materials.
This configuration of a circular Dirac fermion resonator embedded in a constriction is therefore expected to exhibit strong signatures of WGMs in electrical transport.

We now provide an experimental demonstration of this theoretical prediction, by generating a tunable p-n junction in a graphene device \cite{Brun-PrB-2019} using the polarized metallic tip of an atomic force microscope \cite{Eriksson-1996}. Scanning gate microscopy (SGM) consists in scanning a polarized tip in the vicinity of a mesoscopic device, while recording its conductance. It has initially been developed to study mesoscopic phenomena in III-V semiconductor heterostructures \cite{Topinka-2000, Topinka-2001}, in which an insulating layer prevents tunneling from a STM tip to the surface, and has been successfully been extended to the study of transport in graphene devices \cite{Connolly-2011,Pascher-2012, Garcia-2013}.
The sample (see methods), consists of a h-BN-encapsulated monolayer graphene flake, in which a 250~nm-wide constriction is defined by etching \cite{Terres-2016}, consistent with the geometry defined in the simulation.
We first place the tip 90~nm above the constriction center, as sketched in Fig.~\ref{fig2}a, and record the device resistance $R$ while varying the tip and backgate voltages $V_{tip}$ and $V_{bg}$, the experimental knobs linearly related to $\Delta n_{max}$ and $n_{bulk}$ through the relations $\Delta n_{max} = C_{tip} (V_{tip}- V_{tip}^0)$ and $n_{bulk} = C_{bg} (V_{bg}- V_{bg}^0)$, where $C_{tip}$ is the tip-graphene capacitance, $C_{bg}$ the back-gate capacitance, $V_{tip}^0$ and $V_{bg}^0$ are the charge neutrality points when sweeping the tip and back-gate voltages respectively. Details about the determination of the capacitances and neutrality voltages are provided in supplementary materials.
The result, measured at low temperature (4.2 K), is shown in Fig.~\ref{fig2}c, where different regions can be identified (NN'N, NPN, or PP'P), depending on the type of charge carriers below the tip, and in the regions away from the constriction.
When the tip voltage compensates its work-function, i.e. $\Delta n_{max} = 0$, one recovers an unpertubed behavior of the sample and the resistance shows a single maximum at the Dirac point, as shown in Fig.~\ref{fig2}e.
Conversely, quasi-periodic oscillations are visible in the n-p-n configuration, similar to the corresponding simulation of Fig.~\ref{fig2}b. This corresponds to a bulk occupied by electrons and holes in the tip-perturbed region.
For comparison, we plot in Fig.~\ref{fig2}f the resistance as a function of $n_{bulk}$ for a tip voltage of -11~V, corresponding to $\Delta n_{max}$ = $-6.5 \times 10^{15} \rm m^{-2}$, similar to the calculation presented in Fig.~\ref{fig2}d. The qualitative agreement in the evolution of the oscillations in Fig.~\ref{fig2}d and \ref{fig2}f in the range between 2$\times 10^{15}$ m$^{-2}$ and 
5$\times 10^{15}$ m$^{-2}$, and in particular in the positions of the seven minima marked by colored dots in Fig.~\ref{fig2}d, represent a first hint that the experimental resistance oscillations correspond to the internal WGMs of the p-n island.

However, the patterns of Figs.~\ref{fig2}b and \ref{fig2}c are reminiscent of FP oscillations observed in straight gate-defined n-p-n junctions \cite{Young-2009,Rickhaus-2013}. In a recent work \cite{Brun-2020}, we showed that when the tip is centered above the constriction, these tip-induced oscillations exhibit similarities with FP oscillations. In particular their contrast can be controlled by the potential smoothness, that governs the quasi-confinement of Dirac fermions.
Indeed, FP oscillations share a common ground with WGMs', both arising due to coherent superposition of waves in a confined geometry. However, as indicated above, FP are actually a subclass of WGMs in circular or elliptic cavities, corresponding to the fundamental angular momentum m. 
The main question at this point is therefore whether signatures of higher order symmetry WGMs can be clearly evidenced and controlled when charge carriers are transmitted through the circular cavity, making new quantum whisperitronic applications possible.

\begin{figure}
\begin{center}
 \includegraphics[width = 0.45 \textwidth]{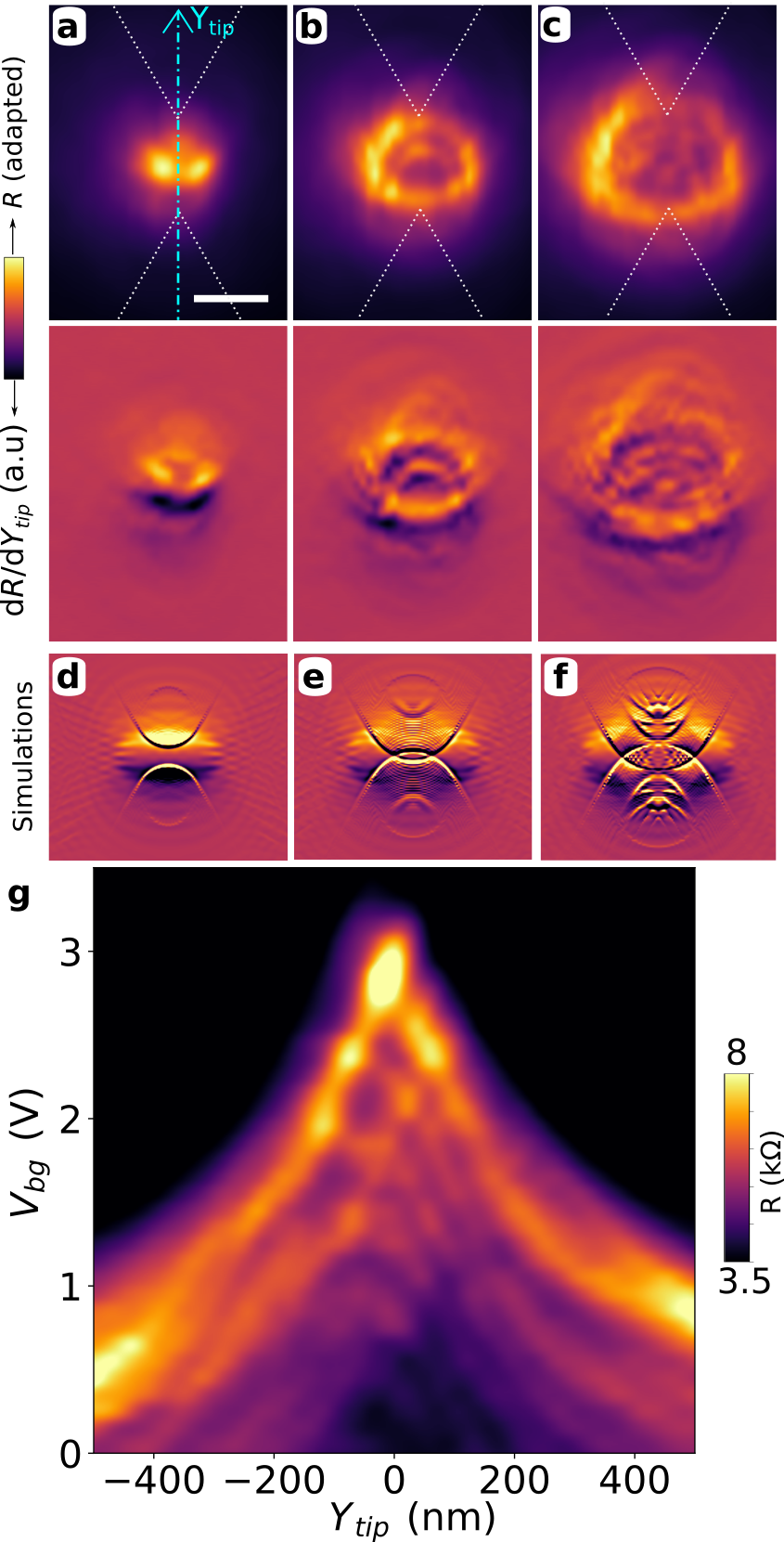}
\caption{\label{fig3} \textbf{Experiment: Scanning gate microscopy images and spectroscopy.} \textbf{a-c}, Up: SGM images, i.e. device resistance as a function of $X_{tip}$ and $Y_{tip}$, for $V_{bg}$ = +1.5~V, $Z_{tip}$ = 90~nm and $V_{tip}$ = -6~V (\textbf{a}), -8~V (\textbf{b}) and -10~V (\textbf{c}). Scale bar: 200~nm. Resistance colorscale has been adapted for each image. Down: derivative of the same SGM images versus $Y_{tip}$ (vertical direction).
\textbf{d-f}, Simulated differentiated SGM maps, for 3 different tip-induced density changes corresponding to the experimental conditions of {a-c} respectively.
\textbf{g}, Measured resistance as a function of $V_{bg}$ and tip position along the blue dash-dotted line in Fig.~\ref{fig3}a, for a tip voltage $V_{tip}$ = -14~V and a tip-to-graphene distance $Z_{tip}$ = 110~nm. 
}
\end{center}
\end{figure}

To reveal the wealth of oscillations arising from internal resonances of the tip-induced p-n island, one needs to off-center the tip with respect to the constriction. Intuitively, gradually overlapping the circular cavity with the etched triangle-shaped edges of the constriction will strongly alter high angular momentum radial resonances as it will break circular symmetry.  
We first scan the polarized tip around the constriction, and map out resistance as a function of tip position for different $V_{tip}$ and $V_{bg}$, yielding the SGM maps shown in Fig.~\ref{fig3}a to \ref{fig3}c (top panels).
We showed in Ref.~\cite{Brun-PrB-2019} that the resistance maps obtained in this type of experiment reflect the average Dirac fermions flow through the p-n island.
Here we go one step further with higher resolution SGM maps and spectroscopies, showing that displacing the p-n island with respect to the constriction allows to probe its internal LDOS resonances. 
In the bottom panels of Figs.~\ref{fig3}a to \ref{fig3}c, we plot the derivative of the resistance versus y-direction (transverse to transport axis) for each of the top panels maps. The differentiated maps better highlight oscillations discussed in the remaining part of the paper.
These maps indeed exhibit arcs centered on the constriction apexes, that are limited to the constriction area. Such arcs are fundamentally different from the usually observed concentric circles, ubiquitous in SGM of quantum dots \cite{Pioda-2004,Schnez-2010,Walkup-2020} or disordered systems\cite{Bleszynski-2007,Pascher-2012,Liu-prB-2015}. The latter circles correspond to Coulomb blockade peaks forming when the confined resonant modes associated with local cavities induced e.g. by disorder are raised locally to the Fermi level by the gating action of the tip. 
Contrary to Coulomb blockade-related circles, the arcs visible in Figs.~\ref{fig3}b-c can be reproduced by single particle tight-binding simulations, as shown in Fig.~\ref{fig3}d-f where we present the simulated SGM maps differentiated versus $Y_{tip}$. This therefore rules out a possible Coulombic origin.

To clarify the origin of SGM arcs, we scan the tip along a line transverse to transport axis (blue dash-dotted line in Fig.~\ref{fig3}a) while varying continuously $V_{bg}$, and plot the resulting resistance map in Fig.~\ref{fig3}g. The overall envelope delineating the high resistance region in the lower part of Fig.~\ref{fig3}g (corresponding to n-p-n junction) reflects the tip-induced potential, and can be fitted to evaluate it accurately (see Refs.\cite{Brun-PrB-2019,Brun-2020} and section 6 of the supplementary materials). 
Within this high resistance region, local resistance maxima undergo the same Lorentzian evolution, with a lateral offset with respect to the envelope,
and lead to higher resistance spots at their crossing points. 
These successive maxima correspond to the interference observed in Fig.~\ref{fig2}f and the SGM arcs in Fig.~\ref{fig3}b-c. 
Figure~\ref{fig3}g reveals that these modes are brought to lower bulk energy (i.e. higher energy in the island) when displacing laterally the tip 
with respect to the constriction center, i.e. when the p-n island overlaps with the constriction side.

\begin{figure*}
\begin{center}
 \includegraphics[width = 1.0 \textwidth]{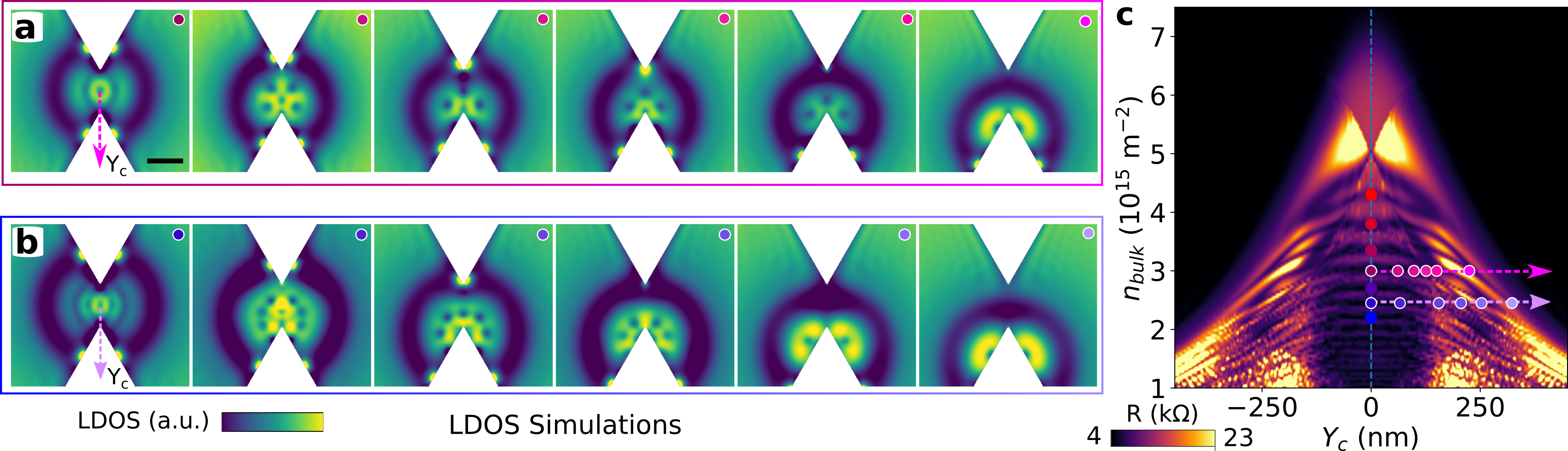}
\end{center}

\caption{\label{fig4} \textbf{Selective transmission of WGMs by lateral displacement:} \textbf{a}, LDOS calculated for the same tip potential as in Fig.~\ref{fig2}, for different tip positions along y-axis
as marked in (c). Scale bar: 200~nm. 
\textbf{b} Same data calculated for a lower bulk energy (i.e. higher energy -hence smaller Fermi wavelength- in the resonator). Corresponding energies and tip positions are indicated by
colored dots in c. \textbf{c} Calculated resistance as a function of energy and lateral p-n island displacement $Y_{c}$, for the same potential as in Figs.~\ref{fig1}d and \ref{fig2}d and \ref{fig4}a-b.
}

\end{figure*}

To understand this observation, we perform additional tight-binding simulations and show the evolution of the LDOS in the p-n island when displacing it along the
axis transverse to the constriction axis in Fig.~\ref{fig4}a and \ref{fig4}b, for two different energies.
From these maps, the role of the constriction on the resonator WGMs is clarified: bringing the tip-induced p-n junction above the etched region reduces the available 
area for the carriers in the p-n island, and consequently promotes lower angular momentum modes in the resonator.
Indeed, depending on the starting eigen-mode, a well-defined number of local LDOS minima are clearly visible as soon as the p-n island is off-centered
by a few tens of nanometers, as visible in second panels of Figs.~\ref{fig4}a and \ref{fig4}b.
This number of ``holes'' is then reduced one by one as the p-n island is brought away from the constriction center (see supplementary movies SM~1-2).
This illustrates that WGMs are morphology-dependent resonances, that can be modified by changing the resonator geometry.

Finally, we present in Fig.~\ref{fig4}c the calculated resistance as a function of density and tip position along y-axis $Y_c$. Though richer details are present compared to the experimental data of Fig.~\ref{fig3}g, the average evolution of the calculated resistance is consistent with the measurement, indicating that the different resonant modes energies are brought to lower bulk density (hence higher holes density in the p-n island) when the tip-induced potential overlaps with the etched region. 

\subsection*{Discussion}

Our results demonstrate two different ways to selectively address the different electrical WGMs of the Dirac fermion resonator:\\
(i) Playing independently on $V_{tip}$ and $V_{bg}$ allows to clearly select the different WGMs, with strong fingerprints in the current through the p-n island. This contrasts with STM experiments, where the tip bias not only changes the resonator shape but also affects the injection energy of charge carriers, leading to a spurious replication of the WGMs at the tip bias energy \cite{Zhao-2015}. Here the injection energy can be tuned completely independently since the tip and the graphene plane are only capacitively connected, so that the intrinsic behavior of WGMs can be more readily accessed.\\
(ii) Displacing the p-n island laterally with respect to the constriction also allows to selectively address the different WGMs. This is a promising configuration since it is analogous to the configuration used in high-quality factor opto-mechanical resonators, where the readout line is side-coupled to the WGMs of the resonator and allows for ultra-fast and sensitive readout of the resonator-state \cite{Rosenblum-2015}.

We would like to emphasize that in the studied geometry, both radial (``Fabry-P\'erot'' like) and higher angular momentum WGM resonances contribute to transport. Engineering further the p-n junction potential and increasing the quality and geometry  of the constriction may allow in the future to more accurately select the different modes contributions to transport.
Additional possibilities for coupling WGMs and a transport channel, as well as for addressing individual WGMs, can also be envisioned. 
For example, in analogy with the geometry of some optical WGM devices, we anticipate that lateral tunnel-coupling of a propagating one-dimensional graphene channel with a circular resonating cavity would provide an even more sensitive and tunable platform to probe individual WGMs. 
It would indeed combine a less invasive approach with the extreme tunability of coupling offered by the tunnel barrier. 
Finally, the high quality factor required to improve the sensitivity of such devices can readily be engineered by tuning the potential smoothness, as reported in Ref.\cite{Brun-2020}.
Opening the way towards such perspectives, this work bridges the recently discovered WGMs of graphene p-n islands with the field of Dirac fermion optics, heralding the advent of relativistic  whisperitronics, a promising field for the engineering of disruptive quantum devices.

 \bibliography{bib_tot}

\section*{Acknowledgments}
We thank Prof. J.A. Stroscio for enthusiastic support and valuable inputs. B.B thanks C. Groth for valuable help regarding Kwant simulations. Computational resources have been provided by the Consortium des \'Equipements de Calcul Intensif (C\'ECI), funded by the Fonds de la Recherche Scientifique de Belgique (F.R.S.-FNRS) under Grant No. 2.5020.11 and by the Walloon Region.
The present research was funded by the F\'ed\'eration Wallonie-Bruxelles through the ARC Grants No. 16/21-077 and No. 21/26-116
and from the European Union's Horizon 2020 Research and Innovation program (Core 3 No. 881603).
B.B. (research assistant), N.M. (FRIA fellowship), B.H. (research associate and PDR No. T.0105.21), V.-H.N. and J.-C.C. (PDR No. T.1077.15 and ERA-Net No. R.50.07.18.F) 
acknowledge financial support from the F.R.S.-FNRS of Belgium.
Support by the Helmholtz Nanoelectronic Facility (HNF), the EU ITN SPINOGRAPH and the DFG (SPP-1459) is gratefully acknowledged. This work was also supported by the FLAG-ERA grant TATTOOS, by the Deutsche Forschungsgemeinschaft (DFG, German Research Foundation) - 437214324.
K.W. and T.T. acknowledge support from the Elemental Strategy Initiative conducted by the MEXT, Japan ,Grant Number JPMXP0112101001, JSPS KAKENHI Grant Numbers JP20H00354 and the CREST(JPMJCR15F3), JST.

%
%

\paragraph*{Supporting informations:\\}

The Supporting Information is available free of charge on the ACS Publications website at DOI: XXXX. 
Supporting Information include figures S1 to S6 and a text detailing sample and measurement methods, individual contributions of the WGMs to the total LDOS, labelling the WGMs in transport, scaling of the tight-binding model, tip-induced potential, simulation of SGM images and effect of npn junction shape on WGM resonances. Two Supplementary Movies (SM1 and SM2) show the evolution of LDOS for the 4th and 6th resonant mode when displacing the tip perpendicular to transport axis.

\paragraph*{Figures.}

\newpage


\end{document}